\documentclass{ws-ijmpa}

\newcommand{\beq}{\begin{equation}}
\newcommand{\eeq}{\end{equation}}

\begin{document}

\markboth{Ale\v{s} Ciepl\'{y}}{}

\catchline{}{}{}{}{}

\title{BARYONIC DECAYS OF CHARMONIUM - A WINDOW ON INTERNAL BARYON
STRUCTURE}

\author{A. Ciepl\'{y}} \address{Institute of Nuclear Physics,
CZ-250 68 \v{R}e\v{z}, Czechia; \\
Institute of High Energy Physics, P.O.Box 918, Beijing 100049,
China \\ cieply@ujf.cas.cz}



\maketitle


\begin{abstract}
The baryonic decays of $J/ \psi$ provide a new way to study the
internal structure of baryons. A simple diquark model applied to
the calculation of the $\bar{B}B$ decay cross-sections is compared
with the ordinary constituent quark model. Various models also give
different predictions for the rates involving the $N^{\ast}(1440)$
resonance in the final state.
\end{abstract}

\keywords{decays of $J/ \psi$; hadron production; diquarks;
Roper resonance}


\section{Introduction}
\label{sec:int}

In spite of a long history of baryon spectroscopy there are still
many questions without clear answers. We know that the baryons are
composed of three valence quarks, sea quarks and gluons but we are
not certain if the valence quarks have constituent or current
nature, if they cluster into diquarks or are well separated from
each other etc. For a long time we have also missed direct
sources of information on the properties of nucleon excitation
states and could build our knowledge about them almost entirely on
results from partial wave analysis of $\pi N$ scattering data. The
situation is changing dramatically with new experimental data
coming from facilities such as CEBAF at JLAB, ELSA in Bonn, GRAAL
in Grenoble or from BEPC in Beijing. Especially the recent (and still
not conclusively confirmed) discovery of pentaquark state is adding
a flavour to the on-going discussion on the baryon and possibly
multiquark internal structures.

A long-standing problem in $N^{\ast}$ physics is about the nature
of the Roper resonance $N^{\ast}(1440)$ which is considered to be
the first radial excitation state of the nucleon. However, various
quark models have difficulty to explain its mass and
electromagnetic coupling, so it was suggested that it may be a
gluonic excitation of the nucleon, a hybrid baryon. The $J/ \psi$
decays into baryon-antibaryon states discussed in the present
contribution provide a novel way to probe the internal structure
of baryons, their resonances and maybe the exotic states.

\section{The models}
\label{sec:mod}

The $J/ \psi$ decay cross sections for various $\bar{B}B$
final states were calculated in Refs.\cite{01zpp,02pcz} by
using a simple quark model. The authors assumed that the decays
proceed in two steps. First, the $\bar{c}c$ pair annihilates in
three gluons and each gluon forms the quark-antiquark pair. Then,
the three quarks and antiquarks combine to form the final state
baryon and antibaryon. The lowest order of perturbative QCD
and constituent quark model wave functions were used to calculate
the relative branching ratios and the angular distributions of
the specific final states.

More recently\cite{03cz} we used the diquark model to look into the
possibility of forming the final state baryons from the $\bar{q}q$
(generated in $J/ \psi$ decay) and diquark-antidiquark
$\bar{D}D$ pairs (created as vacuum excitation). The model
resembles the standard quark-pair-creation model\cite{73yop} used
extensively to describe the mechanism of hadronic decays. In this
model, the transition matrix elements for charmonium decay into
$\bar{B}B$ states are expressed (see Ref.\cite{03cz} for details)
as products of three factors: the amplitude of creating a specific
$\bar{q}q$ pair, the spin-flavour overlap amplitude following the
standard SU(3) quark-diquark decomposition of the baryonic states,
and the space integral involving the intrinsic wave functions of
the baryon and the intermediate $\bar{q}q$ states.

Finally, we refer to the model suggested in Ref.\cite{04pcz} which
uses the Roper resonance $N^{\ast}(1440)$ as a mixture of the standard
three quark state $\mid qqq, 2s \rangle$ and the gluon flavoured
hybrid state $\mid qqqg \rangle$. In this model one of the three gluons
created in the $\bar{c}c$ annihilation may become a constituent of the Roper
resonance formed in the final state. At the same time one of the three
$\bar{q}q$ pairs combining into the $\bar{B}B$ final state is created
as vacuum excitation.

\begin{figure}[h]
\centerline{\psfig{file=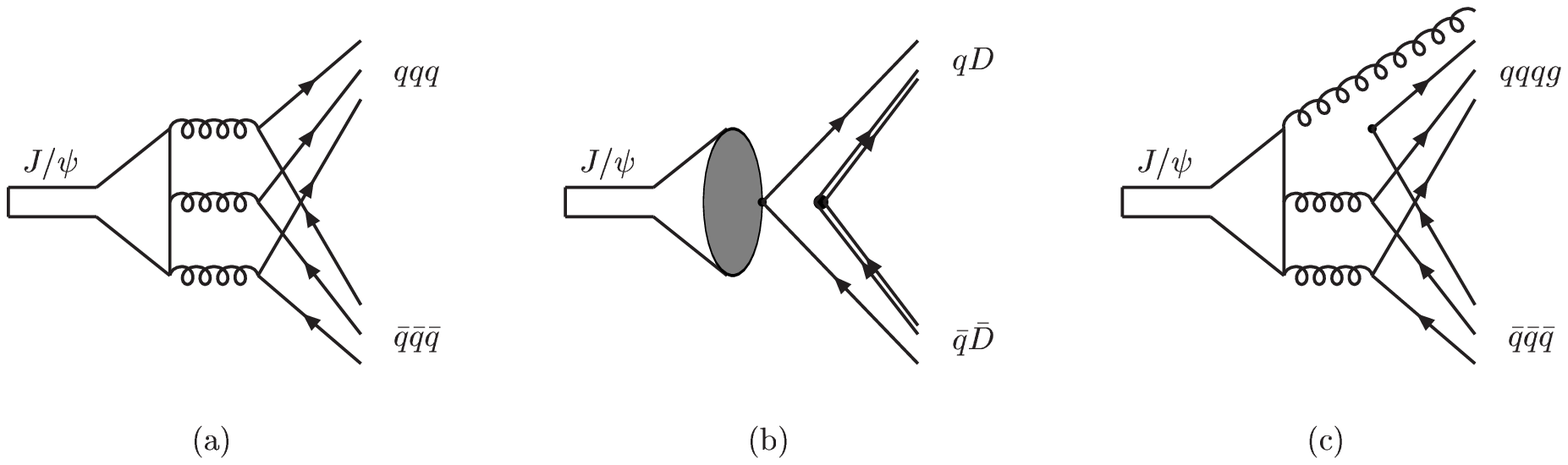,width=13cm,height=4.5cm}}
\caption{A schematic illustration of the discussed models: (a) 
the standard $\mid qqq \rangle$ quark model, (b)~the~$\mid Dq \rangle$ 
diquark model, (c) the hybrid $\mid qqqg \rangle$ model.}
\end{figure}

In the calculations simple gaussian distributions were used for the
intrinsic baryon wave functions (and for the intermediate $\bar{q}q$
states appearing in the diquark model). Their harmonic oscillator parameters
characterize the sizes of the relevant interaction
and we varied them to fit the experimental data. Following Ref.\cite{02pcz}
both the nonrelativistic and relativistic treatments are considered in our
calculations. In the later one the space integral includes the
jacobian due to Lorentz transformation from the baryon CMS to the laboratory
($J/ \psi$ at rest) system and the internal quark-diquark momenta are
transformed appropriately as well\cite{02pcz,03cz}.

\section{Results and discussion}
\label{sec:res}

In all models quoted above the decay cross-section for
$J/ \psi \longrightarrow \bar{B}B$ is constructed from the
corresponding transition amplitudes and is expressed as
\beq \frac{d\Gamma(J/ \psi
\rightarrow \bar{B}B)}{d\Omega} = N_{\bar{B}B}\, (1+a_{B}\cos^{2}
\theta )\;\;\;\; .
\eeq
The constant $N_{\bar{B}B}$ is directly related to the given
experimental branching ratio and we used $N_{\bar{p}p}$ to fix
the overall normalization of the computed rates. The shape of
the angular distribution, i.e. the value of $a_{B}$, was determined
by the parameters $\alpha$ (for the intermediate $\bar{q}q$ state)
and $\beta$ (for the baryon wave function) that characterize the
intrinsic quark distributions.

The results of our calculation performed with the diquark model\cite{03cz}
are presented in the Table~\ref{tab:rate} for both the nonrelativistic
and the relativistic approaches. We show the relative decay rates
$\Gamma_{\bar{B}B}/ \Gamma_{\bar{p}p}$, where $\Gamma_{\bar{B}B}=4\pi
N_{\bar{B}B}(1+a_{B}/3)$, and the angular distribution
coefficients $a_{B}$ in comparison with the available experimental
data. As the reliable data are limited only to the $\bar{p}p$
channel we further assumed $\alpha=\beta$ and made an one parameter
fit to the measured value of $a_{p}$. The results shown in the
Table~\ref{tab:rate} were obtained for $\alpha=\beta=0.4$ GeV and
for $\alpha=\beta=0.22$ GeV in the nonrelativistic and the
relativistic cases, respectively. Although the assumption of using
the same size parameters for both the intermediate $\bar{q}q$
state and for the space distribution of baryon clusters may not be
sound the fitted values compare well with those used in other
quark models\cite{02pcz,ABS96}.

\begin{table}[h]
\tbl{The computed characteristics $\Gamma_{\bar{B}B}/
\Gamma_{\bar{p}p}$ and $a_{B}$ of the $J/ \psi \rightarrow
\bar{B}B$ decay rates.}
{\begin{tabular}{|c|cc|cc|cc|}
\hline\hline
 & \multicolumn{2}{c|}{nonrelativistic case} & \multicolumn{2}{c|}{relativistic case}
 & \multicolumn{2}{c|}{experiment} \\
$\bar{B}B$ state & $\Gamma_{\bar{B}B}/\Gamma_{\bar{p}p}$ & $a_{B}$
& $\Gamma_{\bar{B}B}/\Gamma_{\bar{p}p}$ & $a_{B}$
& $\Gamma_{\bar{B}B}/\Gamma_{\bar{p}p}$ & $a_{B}$ \\
\hline
 $\bar{p}p$ & $1.00$ & $0.64$ & $1.00$ & $0.60$ & $1.00$ & $0.61(11)$\cite{DM2} \\
 $\bar{p}N^{\ast}$ & $0.90$ & $0.53$ & $0.76$ & $0.68$ & $---$ & $---$  \\
 $\bar{N^{\ast}}N^{\ast}$ & $1.05$ & $0.18$ & $0.99$ & $0.32$ & $---$ & $---$  \\
 $\bar{\Lambda}\Lambda$ & $0.56$ & $0.50$ & $0.54$ & $0.54$ & $0.61(9)$\cite{PDG} & $0.62(22)$\cite{DM2}    \\
 $\bar{\Sigma^{0}}\Sigma^{0}$ & $0.41$ & $0.50$ & $0.39$ & $0.56$ & $0.60(11)$\cite{PDG} & $0.22(31)$\cite{DM2}  \\
\hline\hline 
\end{tabular}\label{tab:rate}}
\end{table}

The results are compatible with available experimental data on
$\bar pp$, $\bar\Lambda\Lambda$ and $\bar\Sigma^0\Sigma^0$
channels within two standard deviations. However the predicted
results for the $\bar pN^*(1440)$ and $\bar N^*N^*$ are quite
different from those given by the simple $\mid qqq \rangle$ quark
model\cite{02pcz} and for the pure $\mid qqqg \rangle$ hybrid
state\cite{04pcz}. The comparison of all three models is given in the
Table~\ref{tab:roper} where our diquark model is labeled by
$\mid Dq \rangle$. The authors of Refs.\cite{02pcz} and \cite{04pcz}
varied the model parameters to get the results within the limits shown
in the Table~\ref{tab:roper}. Future experimental results on the
$N^*(1440)$ channels will be helpful for examining various model
predictions and to improve our understanding of internal quark
structure of these baryons. If the experimental $N^{*}$ production
rates turn out to be much lower than the quark (and diquark) model
predictions suggest a large component of the hybrid $\mid qqqg \rangle$
state could contribute to the $N^{*}$ internal structure.
However, the simple models considered in this contribution cannot rule
out conclusively some other mechanisms playing a role.

\begin{table}[h]
\tbl{The comparison of the results obtained for the $N^{\ast}(1440)$
channels within different models. Only the relativistic case is shown.}
{\begin{tabular}{|c|ccc|ccc|}
\hline\hline
 & \multicolumn{3}{c|}{$\Gamma_{\bar{B}B}/\Gamma_{\bar{p}p}$}
 & \multicolumn{3}{c|}{$a_{B}$} \\
$\bar{B}B$ state & $\mid Dq \rangle$ & $\mid qqq \rangle$
                 & $\mid qqqg \rangle$
                 & $\mid Dq \rangle$ & $\mid qqq \rangle$
                 & $\mid qqqg \rangle$ \\
\hline
 $\bar{p}N^{\ast}$        & $0.76$ & $2.0\: - \:4.5$ & $<\:0.02$
                          & $0.68$ & $0.22$ - $0.70$ & $0.42\: -\: 0.57$ \\
 $\bar{N^{\ast}}N^{\ast}$ & $0.99$ & $3.2\: -\: 22.0$ & $<\:0.002$
                          & $0.32$ & $0.06\: -\: 0.08$ & $-(0.1\: -\: 0.9$) \\
\hline\hline
\end{tabular}\label{tab:roper}}
\end{table}

\section{Conclusions}

Baryonic $J/ \psi$ decays offer a novel tool to study the internal
quark structure of baryons. Our diquark model calculation gives an
alternative to more common quark models with different predictions
for the formation rates of the $\bar{B}B$ final states. Specifically,
the results obtained for the channels involving the $N^*(1440)$ Roper
resonance depend significantly on the employed model. The calculations
also show the importance of relativistic treatment for the distribution
of baryon constituents. Future experimental data may provide a check on
these different pictures for baryons.


\section*{Acknowledgements}

This research was partially supported by the Grant Agency of
the Czech Republic (grant No. 202/00/1667), CAS
Knowledge Innovation Project (KJCX2-SW-N02) and the National
Natural Science Foundation of China. The author acknowledges the
hospitality of the IHEP in Beijing.


\end{document}